# Deep Semantic Architecture with discriminative feature visualization for neuroimage analysis

**Arna Ghosh**
Integrated Program in Neuroscience
McGill University
Montréal, QC H3A 0G4. Canada
`arna.ghosh@mail.mcgill.ca`

**Fabien Dal Maso**
École de kinésiologie et des
sciences de l'activité physique
Université de Montréal
Montréal, QC H3T 1J4. Canada
`fabien.dal.maso@umontreal.ca`

**Marc Roig**
School of Physical and
Occupational Therapy
McGill University
Montréal, QC H3A 0G4. Canada.
`marc.roigpull@mcgill.ca`

**Georgios D Mitsis**
Department of Bioengineering
McGill University
Montréal, QC H3A 0G4. Canada.
`georgios.mitsis@mcgill.ca`

**Marie-Hélène Boudrias**
School of Physical and
Occupational Therapy
McGill University
Montréal, QC H3A 0G4. Canada.
`mh.boudrias@mcgill.ca`

## Abstract

Neuroimaging data analysis often involves *a-priori* selection of data features to study the underlying neural activity. Since this could lead to sub-optimal feature selection and thereby prevent the detection of subtle patterns in neural activity, data-driven methods have recently gained popularity for optimizing neuroimaging data analysis pipelines and thereby, improving our understanding of neural mechanisms. In this context, we developed a deep convolutional architecture that can identify discriminating patterns in neuroimaging data and applied it to electroencephalography (EEG) recordings collected from 25 subjects performing a hand motor task before and after a rest period or a bout of exercise. The deep network was trained to classify subjects into exercise and control groups based on differences in their EEG signals. Subsequently, we developed a novel method termed the cue-combination for Class Activation Map (ccCAM), which enabled us to identify discriminating spatio-temporal features within definite frequency bands (23–33 Hz) and assess the effects of exercise on the brain. Additionally, the proposed architecture allowed the visualization of the differences in the propagation of underlying neural activity across the cortex between the two groups, for the first time in our knowledge. Our results demonstrate the feasibility of using deep network architectures for neuroimaging analysis in different contexts such as, for the identification of robust brain biomarkers to better characterize and potentially treat neurological disorders.

# 1 Introduction

Skilled motor practice facilitates the formation of an internal model of movement, which may be later used to anticipate task specific requirements. These internal models are more susceptible to alterations during and immediately following practice and become less susceptible to alterations over time, a process called consolidation [5, 23]. A single bout of cardiovascular exercise, performed in close temporal proximity to a session of skill practice, has shown to facilitate motor memory consolidation [24]. Several potential mechanisms underlying the time-dependent effects induced by acute exercise on motor memory consolidation have been identified, such as increased availability of neurochemicals [28] and increased cortico-spinal excitability [19]. However, the distinct contribution of specific brain areas and the precise neurophysiological mechanisms underlying the positive effects of acute cardiovascular exercise on motor memory consolidation remain largely unknown.

Electroencephalography (EEG) is a popular technique used to study the electrical activity from different brain areas. The EEG signal arises from synchronized postsynaptic potentials of neurons that generate electrophysiological oscillations in different frequency bands. During movement, the EEG signal power spectrum within the alpha (8–12 Hz) and beta (15–29 Hz) range decreases in amplitude and this is thought to reflect increased excitability of neurons in sensorimotor areas [7, 18, 20, 25]. This phenomenon is termed Event-Related Desynchronization (ERD). Alpha- and beta-band ERD have been shown to be modulated during motor skill learning in various EEG studies [4, 14, 34]. There is converging evidence suggesting an association of cortical oscillations in the motor cortex with neuroplasticity events underlying motor memory consolidation [4, 22]. In this context, our aim was to study the add-on effects of exercise on motor learning in terms of modulation of EEG-based ERD.

Many neuroimaging studies, including EEG ones, rely on the *a-priori* selection of features from the recorded time-series. This could lead to sub-optimal feature selection and could eventually prevent the detection of subtle discriminative patterns in the data. Alternatively, data-driven approaches such as deep learning allow discovery of the optimal discriminative features in a given dataset. Convolutional Neural Networks (CNNs) and Recurrent Neural Networks (RNNs) have been applied to computer vision and speech processing datasets [17, 12, 32, 15] with great success. They have also been used successfully in the neuroimaging domain to learn feature representations for Magnetic Resonance Image segmentations [21] and EEG data decoding [2, 26, 30] among others. Most studies using CNNs for EEG have been restricted to the classification of EEG data segments into known categories. However, the usefulness of CNNs to improve our understanding of the underlying neural bases is less straightforward, primarily due to the difficulty into visualizing and interpreting the feature space learnt by the CNN.

Our work addresses existing caveats in applying deep learning architectures, such as CNNs, to analyzing EEG data by means of three novel contributions –

1. We used two parallel feature extraction streams to discover informative features from EEG data before and after an intervention and subsequently characterize the modulatory effect on these derived features rather than on the raw EEG data itself;
2. We incorporated a subject prediction adversary component in the network architecture to learn subject-invariant, group-related features instead of subject-specific features;
3. We developed a novel method, termed cue-combination for Class Activation Map (ccCAM), to visualize the features extracted by the CNN after training

We used this CNN-based deep network architecture to identify exercise-induced changes in neural activity from EEG signals recorded during an isometric motor task. The training was carried out in a hierarchical structure – first for time-frequency and then for topographical data maps. Visualizing the features after each stage of the training allowed us to identify frequency bands and the corresponding brain regions that were modulated by the add-on effects of acute exercise on motor learning.

The majority of previous related studies have leveraged large-scale datasets consisting of hundreds of subjects for training purposes, which may be a limiting factor for applying powerful deep learning methodologies to data of smaller sample size. Therefore, one of our aims was to develop a method that can be used both for small-scale and large-scale studies. To this end, we added a regularizer that prevented the feature extraction part of the CNN from learning subject-specific features, thus promoting the identification of group-specific features only.

## 2 Dataset

The dataset used in this work consisted of EEG recordings from 25 healthy subjects. The experiment and data collection are detailed elsewhere [8]. Briefly, 25 right-handed healthy subjects were recruited and assigned to the Control (CON, n=13 subjects) or Exercise (EXE, n=12 subjects) groups.

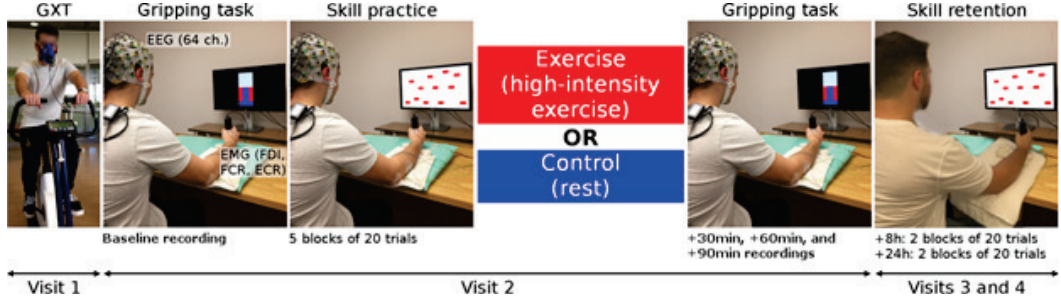

Figure 1: Illustration of Experiment Protocol.

Each subject reported to the laboratory on four occasions as shown in Figure 1. Visit 1 required the participants to go through a Graded Exercise Test (GXT), which was used to determine their cardiorespiratory fitness. Visit 2 was conducted at least 48 hrs after the GXT to avoid potential long-term effects of exercise on memory [3, 13]. EEG recordings were collected at baseline while subjects performed isometric handgrips, which corresponded to 50 repetitions of visually cued isometric handgrips with their dominant right hand using a hand clench dynamometer (Biopac, Goleta, CA, USA). Each contraction was maintained for 3.5 sec at 15% of each participant's maximum voluntary contraction (MVC) and followed by a 3 to 5 sec rest period. The baseline assessment was followed by the practice of a visuo-motor tracking task (skill acquisition), which was used for the calculation of the motor learning score. Participants were then randomly assigned to two groups. The exercise group (EXE) performed a bout of high-intensity interval cycling of 15 min, while the control group (CON) rested on the cycle ergometer for the same amount of time. The same EEG recordings collected at baseline were repeated 30, 60 and 90 min after the exercise or rest period.EEG activity was recorded using a 64-channel ActiCap cap (BrainVision, Munich, Germany) with electrode locations arranged according to the 10–20 international system. Electrical conductive gel was inserted at each electrode site to keep impedances below 5 k$\Omega$. EEG signals were referenced to the FCz electrode and sampled at 2500 Hz.

## 3 Methods

The analysis pipeline was first applied to the time and frequency domain data without incorporating spatial information. Subsequently, it was applied to the data obtained by creating topographical maps corresponding to distribution of activity in specific frequency bands across the cortex. The entire pipeline consisted of 3 segments , i.e.– Preprocessing, CNN training and ccCAM generation.

### 3.1 Time-Frequency (TF) maps

#### 3.1.1 Preprocessing

EEG data preprocessing was similar to that performed in a previous study [8] and was performed using the Brainstorm Matlab toolbox [29]. EEG signals were bandpass-filtered between 0.5 Hz and 55 Hz and average-referenced. Continuous data were visually inspected and portions of signals with muscle or electrical transient artifacts were rejected. Independent component analysis (ICA) was subsequently applied on each dataset (total number of components: 20) and between one and two eye-blink related components were rejected based on their topography and time signatures [9]. The resulting dataset was epoched with respect to the period of time (3.5 sec) corresponding to the appearance of the visual cue that triggered the initiation of the isometric handgrips (n = 50/subject). Finally, each trial was visually inspected and those containing artifacts were manually removed.

Morlet wavelet (wave number = 7) coefficients between 1 to 55 Hz with 1 Hz resolution were extracted to obtain time-frequency decompositions of the EEG data. The time-frequency data for each electrode were consequently normalized with respect to their spectral power before the start of the grip event, as calculated from a window of 0.5 sec. Following this, an average over all trials was calculated in order to obtain a single time-frequency map for each electrode. Further steps were applied on the EEG recording segment corresponding to 0.5–3.5 sec after the appearance of the visual cue, i.e. during the handgrip task, to perform the subsequent analysis.

### 3.1.2 CNN training

The overall CNN architecture that we developed is shown in Figure 2. Following preprocessing of the data, time-frequency maps for each electrode and session – at baseline and 90 min after exercise or a rest period (post-intervention session) – were obtained. The data for each session was then rearranged to form 2D matrices comprising of the frequency spectra for all electrodes at a given time instant $t$. Each matrix had a dimension of $64 \times 55$ (64 electrodes $\times$ 55 frequency bands). For training the network, a pair of matrices was used – the first corresponding to time point $t$ from the baseline session and the second corresponding to the same time point $t$ from the post-intervention session. Each pair was labeled as either exercise or control, depending on the group allocation. Structuring the data in this fashion allowed the network to take into account the inter-subject variability in baseline measures and therefore did not require the experimenter to adopt techniques for normalizing the EEG signal from the post-intervention session with respect to the baseline session. Thus, the network was expected to capture the EEG features that were modulated by the add-on effects of acute exercise.

*Dataset Notation*:- $B$ and $A$ represent the entire data tensor at baseline and post-intervention respectively. Each data tensor consists of data matrices from all 25 subjects and timepoints. For subject $s$, the goal was to classify whether the tuple containing the matrices $B_t^s$ and $A_t^s$ (where $t$ denotes timepoint) belongs to the EXE or CON groups.

To this end, we used a deep convolutional network that was optimized for the task. The network architecture is similar to the one used in [1]. Features from matrices $B_t^s$ and $A_t^s$ were extracted using a network termed the Base CNN. The difference between the obtained feature vectors was passed to a discriminator network, termed the Top NN, to predict the correct group in which each pair belongs to. The schematic view of the architecture is shown in Figure 2 and details of each network's architecture are provided in Tables S1 and S2 in supplementary material respectively. Since the sampling frequency was 2500 Hz and the time period of interest was 3 sec long, $t \in [1, 7500]$ for each subject $s$.

The convolutions performed in the Base CNN were with respect to the frequency domain and not the electrode (sensor) domain. This is because the former was laid out in a semantic order of increasing frequencies, as opposed to the latter, which was not arranged by the spatial locations of the electrodes. Consequently, we expected the features extracted by the Base CNN to be the frequency bands significantly affected by exercise. Therefore, all convolutional filters in the Base CNN were implemented as $1 \times n$ 2D filters, where $n$ is the extent of the filter in the frequency domain. The same holds for the Max-Pooling layers.

Initially, a network that did not include an adversary loss component (Figure S1.a from Supplementary material) was used; however, it was found that this network was able to learn subject-specific features as opposed to subject-invariant, exercise-related features. This is illustrated in Figure S2 (Supplementary material) and Table 1. In most neuroimaging studies, the number of participants scanned is limited, which typically restricts deep networks from learning subject-invariant features. To address this issue, we followed a domain adaptation approach. Specifically, each subject was considered as a separate domain comprising of subject-specific features along with subject-invariant, exercise-related features. Since our goal was to learn features mainly related to the effect of exercise on the consolidation of motor memory, we incorporated the domain confusion strategy [31] to train the network, thus adding the subject discriminator as an adversary (Figure 2 – bottom right). Specifically, we added this network in parallel to the Top NN with similar model capacity (see Table S3 in supplementary material for architecture details).

*Network Architecture Notation*:- The feature extractor operation and parameters of the Base CNN are denoted as $f_{\theta_f}$ and $\theta_f$ respectively, the Top NN feature discrimination operator and its parameters are denoted by $h_{\theta_t}$ and $\theta_t$ respectively, while the subject discrimination operator and its parameters are denoted by $h_{\theta_s}$ and $\theta_s$ respectively. The input tuple is denoted by $x$ and its corresponding group and subject labels by $y_g$ and $y_s$ respectively. We used the Negative Log Likelihood (NLL) loss for each classifier with the Adam optimizer [16] in Torch [6] for training the network. The Subject

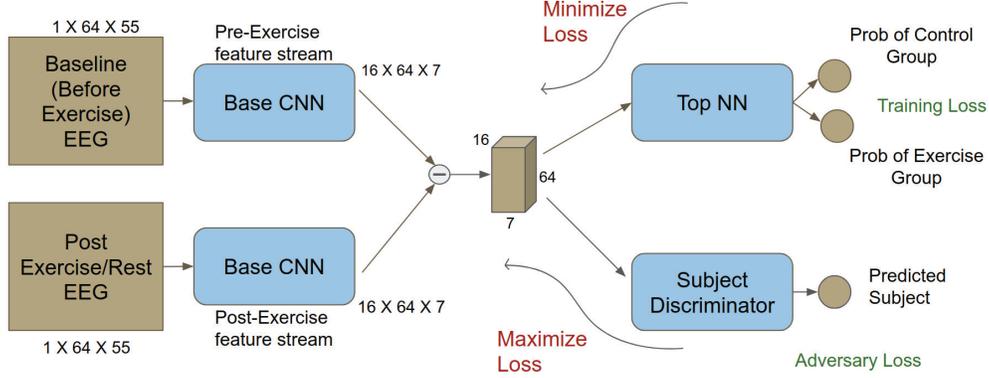

Figure 2: Modified deep network architecture with an adversary component (bottom right) to avoid subject discrimination. The adversary is a novel addition to enable the use of CNNs to smaller-scale neuroimaging studies with a limited number of subjects. Dimensions corresponding to TF maps are shown here.

Discriminator was trained to minimize the subject prediction loss given by –

$$J_s(\theta_s, \theta_f) = -[\sum_{i=1}^{m} log h_{\theta_s}^{(y_s^{(i)})}(f_{\theta_f}(x^{(i)}))] \tag{1}$$

The Top NN was trained to minimize the group prediction loss given by –

$$J_g(\theta_t, \theta_f) = -[\sum_{i=1}^{m} log h_{\theta_t}^{(y_g^{(i)})}(f_{\theta_f}(x^{(i)}))] \tag{2}$$

We trained the feature extractor, Base CNN, in a manner such that the features extracted would be agnostic to the originating subject, therefore, the target distribution for the subject prediction network was a uniform distribution. Hence, we used the domain confusion loss [31] over the gradient reversal layer [11] and used the Kullback-Leibler (KL) divergence from the uniform distribution over 25 classes (25 subjects) as our loss metric. Conclusively, the Base CNN was trained to minimize the loss given by –

$$J_f(\theta_f, \theta_t, \theta_s) = -[\sum_{i=1}^{m} log h_{\theta_t}^{(y_g^{(i)})}(f_{\theta_f}(x^{(i)}))] + \lambda[\sum_{i=1}^{m} KL(U, h_{\theta_s}(f_{\theta_f}(x^{(i)})))] \tag{3}$$

where $KL(P, Q)$ denotes the KL divergence between distributions $P$ & $Q$, $U$ denotes the uniform distribution, $m$ denotes the total number of training examples and $\lambda$ is a hyperparamater that determines the weight for the subject discrimination regularizer. Here, we used a 80-20 split of the data set, whereby 80% was used for training and 20% was used for validation.

### 3.1.3 ccCAM

A major contribution of the present work is the development of a novel method for the visualization of the features that guide the proposed network's decision. Although well-known techniques used in the computer vision literature include the use of Global Average Pooling (GAP) [33] and grad-CAM [27], these methods are not suited for the neuroscience paradigm considered here. For instance, GAP requires averaging the activations of each filter map, i.e. each channel of the extracted feature tensor. This leads to loss of information related to electrode positions, as convolutions were performed only in the frequency domain. Specifically, we applied GAP and grad-CAM to our data and we were unable to obtain adequate classification accuracy ($\approx 56\%$) with a GAP layer in the network. Also, grad-CAM is sensitive to absolute scale of the features in the input data and hence yielded results that were biased towards frequency bands with higher power-values, namely the lower frequency bands (<10 Hz).

Given these limitations in existing analytic methods, we used the linear cue-combination theory used in human perception studies [10] to develop a method that explains the network's decisions. Let us consider for example, a CNN with only 2 channels, i.e. filter maps, in the final feature tensor extracted after convolutions. Each of these filter maps preserve the spatial and/or semantic structure of the input data. Each of these filter maps acts as a "cue" to the network's classifier layers, denoted as $c_1$ and $c_2$. If we denote the desired class label as $y_1$ and assuming $c_1$ and $c_2$ to be independent to each other, we can use Bayes' Theorem to write –

$$P(y_1|c_1,c_2) = \frac{P(c_1,c_2|y_1)P(y_1)}{P(c_1,c_2)} = \frac{P(c_1|y_1)P(c_2|y_1)P(y_1)}{P(c_1)P(c_2)} = \frac{P(y_1|c_1)P(y_1|c_2)}{P(y_1)} \quad (4)$$

If the likelihood for predicting $y_1$ due to cue $c_i$ is Gaussian with mean $\mu_i$ and variance $\sigma_i^2$, the maximum likelihood estimate (MLE) yields the combined cue, denoted by $c^*$, that summarizes the important features on which the network bases its decisions. Therefore, the combined cue, $c^*$, is the desired Class Activation Map (CAM).

$$c^* = \sum_{i=1}^{2} w_i \mu_i \text{ where } w_i = \frac{1/\sigma_i^2}{\sum_{i=1}^{2} 1/\sigma_i^2} \quad (5)$$

Since the network is trained, $\mu_i = c_i$. To calculate the values of $\sigma_i$, we used the NLL loss values. The NLL loss with a cue removed provided an estimate of the $\sigma$ associated with that cue.

$$\begin{aligned}
\epsilon &= -logP(y_1|c_1,c_2) \\
&= -logP(y_1|c_1) - logP(y_1|c_2) + logP(y_1) \text{ [From eq 4]} \\
\epsilon_1 &= \epsilon|_{c_1=0} = -logP(y_1|c_1=0) - logP(y_1|c_2) + logP(y_1) \\
\epsilon_1 - \epsilon &= logP(y_1|c_1) - logP(y_1|c_1=0) \\
&= \frac{\mu_1^2}{2\sigma_1^2} \\
\text{Therefore, } \frac{1}{\sigma_1^2} &= \frac{2(\epsilon_1 - \epsilon)}{\mu_1^2}
\end{aligned} \quad (6)$$

$\sigma_i$ is estimated over the entire dataset as shown in Equation 7.

$$\frac{1}{\sigma_j^2} = \sum_{i=1}^{m} \frac{2[(\epsilon_j - \epsilon)]^{(i)}}{[\mu_j^2]^{(i)}} \quad (7)$$

Using the estimated $\sigma_i$, the CAM corresponding to the correct class for each input was generated. Since in the present case $\mu_i$ corresponds to a 2D matrix, the denominator in Equation 7 was replaced by the mean-squared value of the corresponding matrix. The obtained CAMs were subsequently group-averaged to extract frequency bands that contain features characteristic to each group (CON and EXE).

### 3.2 Topographical maps

Topographical maps were created using the frequency bands obtained from the ccCAM corresponding to the TF-maps. The average power within each frequency band for all electrodes at time point $t$ was used to construct a $64 \times 64$ matrix by projecting the average power value of each electrode to a point corresponding to the its spatial position. Since this procedure yielded a sparse matrix, cubic interpolation was used to obtain a continuous image depicting the distribution of activity within each frequency band over the entire head. A total of three such matrices were packed together to form a $3 \times 64 \times 64$ tensor corresponding to activity maps at times $t$, $t+1$ and $t+2$ respectively. The entire data tensor for a given subject was created by taking non-overlapping time windows. Hence, the total number of tensors for each subject was equal to 2500.

Similar to the analysis of TF-maps, we trained a CNN-based network to classify each data tensor into the CON and EXE groups. Since the inputs were 2D image tensors here, we used 2D convolutional filters in the Base CNN (see Tables S4, S5 and S6 from supplementary material for more details). Following training, ccCAM was applied to obtain CAMs for each subject at each time instant during the task execution.

## 4 Results and Discussion

The results presented here illustrate the differences between the Baseline and 90 min post exercise/rest datasets. The network architecture details for each type of data (TF and Topographical) map are presented in the supplementary material, along with details regarding the chosen hyperparameters.

### 4.1 Time-Frequency maps

We observed that the features extracted by Base CNN, without any subject prediction regularizer, could be used to identify the subject corresponding to any given data tensor. As the subject discriminator regularization was given more weight by increasing $\lambda$, the Base CNN learned to extract features that were agnostic to the originating subject. However for very high $\lambda$ values, the extracted features could not be used to discriminate the EXE and CON groups, suggesting that the Base CNN was unable to learn discriminative features. The loss values obtained post-training for four different values of $\lambda$ are shown in Table 1. The choice of an optimal value for $\lambda$ depends on two factors – group prediction accuracy and subject prediction accuracy. To identify subject-invariant features, we aimed to obtain an optimal value of $\lambda$ that achieved good group prediction accuracy but poor subject prediction accuracy. Consequently, this required a good tradeoff between the two prediction accuracies.

Table 1: Variation of Loss values with $\lambda$ after training network on TF maps.

| $\lambda$ | Group prediction loss (NLL) | Subject prediction loss (NLL) | KL divergence loss from Uniform distribtion |
|---|---|---|---|
| 0 | $\approx 0$ | $\approx 0$ | $\approx 0.3$ |
| 10 | $\approx 0.1$ | $\approx 1.5$ | $\approx 0.07$ |
| 13 | $\approx 0.4$ | $\approx 2.6$ | $\approx 0.004$ |
| 15 | $\approx 0.68$ | $\approx 3.2$ | $\approx 0.0002$ |

According to this procedure, the model corresponding to $\lambda = 13$ was used for ccCAM generation. The average loss over a batch for subject prediction was around 2.6, which roughly predicted the correct subject with a confidence of $\frac{1}{13}$. The group prediction accuracy was **99.984%** (**99.969%** for CON and **100%** for EXE). Hence the extracted features achieved excellent group prediction, while all subjects in the group were predicted with roughly equal probability (CON and EXE consisted of 13 and 12 subjects respectively). The ccCAM obtained is shown in Figure 3.

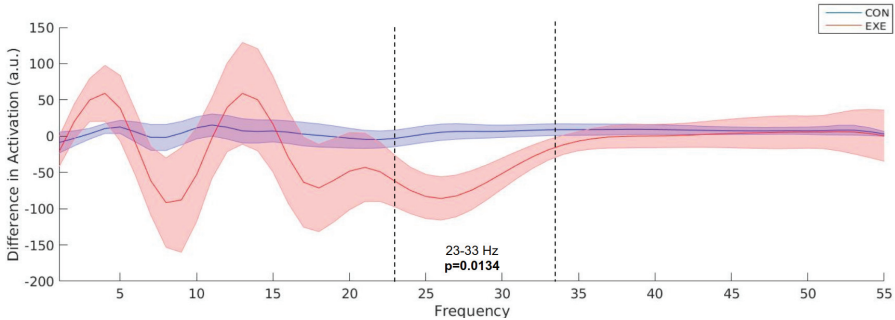

Figure 3: Time-averaged TF map ccCAM averaged over electrodes & subjects showing "discriminative" frequencies.

As one of the main goals of this study was to identify the frequency bands that contained significant information, we calculated the ccCAMs for all timepoints and then group-averaged (averaging across all timepoints and subjects in a group) the maps to get two 2D maps – for the CON and EXE groups. We plotted the average activation within each frequency band in each of these 2D maps to obtain the plots in Figure 3. The bold lines denote the group-mean and the shaded regions span 1 standard error over all subjects in the group. The two plots are significantly different within the band 23–33 Hz. This band lies within the wider beta-band and agrees with findings in [8] where beta-band desynchronization was found to be significantly modulated by exercise. It is important to note that the ccCAM highlights the differences between the 90min and baseline EEG recordings. The negative values in a frequency band indicate that the ERD was smaller after than before the exercise. This also agrees with findings in [8] and implies that decreased neural activity was required to perform the hand-grip task after exercise. The p-value calculated from the ccCAM outputs within this frequency band was equal to **0.021**, while the corresponding p-value from the original time-frequency data tensor was equal to **0.0134**. This suggests that had the band of interest in the previous study [8] been chosen to be 23–33 Hz, instead of the wider beta-band (15–29 Hz), similar, statistically significant inferences would have been drawn.

Topographical maps were created to understand the distribution of the activity within the 23–33 Hz frequency band over the cortex. After training a network to classify into the CON and EXE groups from topographical maps, a classification accuracy of **98.70%** (**98.94%** for CON and **98.43%** for EXE) was obtained for $\lambda = 5$. Generating ccCAMs for the topographical maps revealed the propagation of the discriminative activity across the cortex. A video showing this traveling property of this activity is included in the supplementary material. Some snapshots from the video are shown in Figure 4.

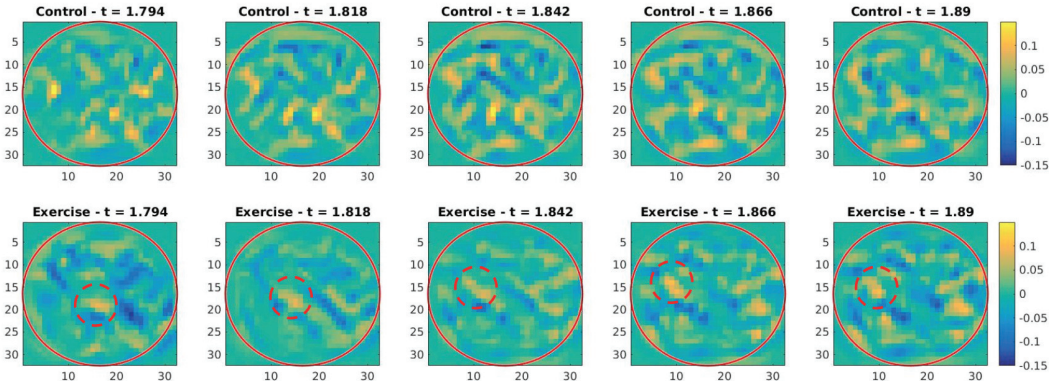

Figure 4: Topographical map ccCAM averaged over subjects in a group showing regions with difference in activity before and 90min after rest/exercise. Top row shows CON group. Bottom row shows EXE group for a brief time window, where a streak of activity can be seen to move across time from the parietal to motor cortex.

To the best of our knowledge, this traveling pattern of activity across the cortex while performing an isometric handgrip has not been demonstrated before. These oscillations could allow us to visualize the neural mechanisms involved in maintaining a constant grip-force output. Further investigation into the correlation of these activities with the observed error signal while performing the task is required to understand these mechanisms more precisely. As expected, differences in the ccCAM of EXE group before and after exercise were higher in magnitude as compared to those in the CON group (see Figure S4 in supplementary material), thus indicating the modulatory effects of an acute bout of high-intensity exercise.

## 5 Conclusion

This work introduces a deep learning architecture for the analysis of EEG data and shows promising results in terms of discriminating the participants that underwent an acute bout of high-intensity exercise/rest in close temporal proximity to performing a motor learning task. Importantly, the proposed novel method enabled us to visualize the features learnt by deep networks such as CNNs, which may in turn yield better interpretation of their classification basis. The results are in general agreement with those reported in a previous study using more standard statistical analysis for a-priori selected features on the same dataset [8], with our analysis revealing a narrower, more-specific frequency band associated with exercise-induced changes. In addition, our method revealed, for the first time, the traveling pattern of cortical activity while subjects were performing isometric handgrips.

Therefore, our approach demonstrates scope of identifying discriminative features in a completely data-driven manner. The proposed method is not restricted to the EEG modality and dataset described here. Hence, it paves the way for applying equivalent deep learning methods to datasets obtained from neuroimaging studies of differing scales and varying modalities (eg. magnetoencephalography – MEG). This, in turn, yields great potential to accelerate research oriented towards identification of neurophysiological changes associated with various neurological disorders and ultimately lead to design of optimized and individualized intervention strategies.

# Deep Semantic Architecture with discriminative feature visualization for neuroimage analysis

**Arna Ghosh**
Integrated Program in Neuroscience
McGill University
Montréal, QC H3A 0G4. Canada
`arna.ghosh@mail.mcgill.ca`

**Fabien Dal Maso**
École de kinésiologie et des
sciences de l'activité physique
Université de Montréal
Montréal, QC H3T 1J4. Canada
`fabien.dal.maso@umontreal.ca`

**Marc Roig**
School of Physical and
Occupational Therapy
McGill University
Montréal, QC H3A 0G4. Canada.
`marc.roigpull@mcgill.ca`

**Georgios D Mitsis**
Department of Bioengineering
McGill University
Montréal, QC H3A 0G4. Canada.
`georgios.mitsis@mcgill.ca`

**Marie-Hélène Boudrias**
School of Physical and
Occupational Therapy
McGill University
Montréal, QC H3A 0G4. Canada.
`mh.boudrias@mcgill.ca`

## Supplementary Material

## 1 Network Architecture

**Notation**:- *Conv* denotes the 2D Spatial Convolutional layer. *ReLU* denotes the Rectified Linear Unit Layer that adds non-linearity to the network. *MaxPool* denotes the 2D Spatial Max Pooling layer. *FullyConn* denotes a Fully Connected layer, also known as the linear layer of the network.

### 1.1 TF maps

| Layer | Type    | Maps and Neurons              | Filter Size |
|-------|---------|-------------------------------|-------------|
| 0     | Input   | $1M \times 64 \times 55N$     | -           |
| 1     | Conv    | $6M \times 64 \times 28N$     | $1\times5$  |
| 2     | ReLU    | $6M \times 64 \times 28N$     | -           |
| 3     | MaxPool | $6M \times 64 \times 14N$     | $1\times2$  |
| 4     | Conv    | $16M \times 64 \times 14N$    | $1\times5$  |
| 5     | ReLU    | $16M \times 64 \times 14N$    | -           |
| 6     | MaxPool | $16M \times 64 \times 7N$     | $1\times2$  |

Table S1: Network architecture used for EEG feature extraction network (Base CNN). The output of the network is a tensor of dimensions $16 \times 64 \times 7$.

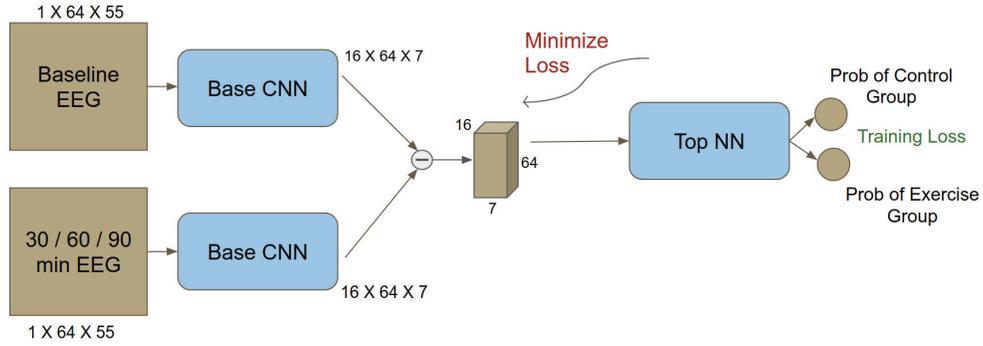
(a) Basic Architecture without adversary

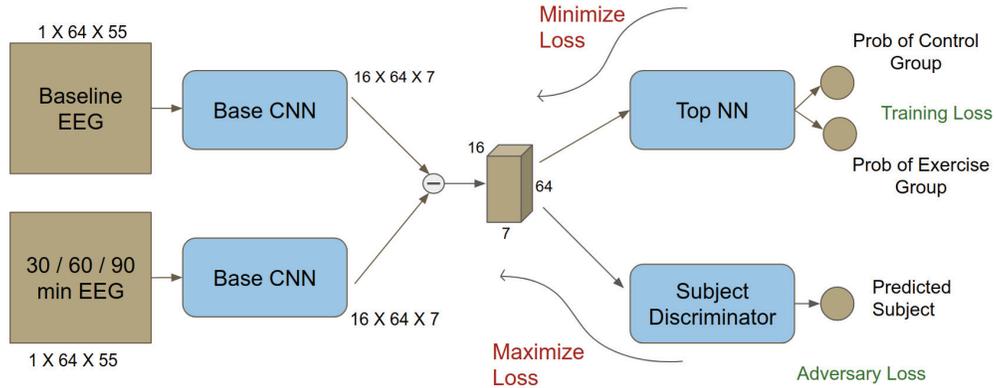
(b) Modified Architecture with adversary to avoid subject discrimination

Figure S1: Deep Network Architecture. The initial choice of architecture (without any adversary) gives good subject prediction accuracy from features extracted by the Base CNN. Therefore, a subject discriminator of roughly the same model capacity as the Top NN is added. The subject discrimination acts as a regularizer while training and avoids the Base CNN from learning subject specific features.

| Layer | Type | Maps and Neurons | Filter Size |
|---|---|---|---|
| 0 | Input | $16M \times 64 \times 7N$ | - |
| 1 | Flatten | $7168N$ | - |
| 2 | Dropout (p=0.5) | - | - |
| 3 | FullyConn | $8N$ | $1\times1$ |
| 4 | ReLU | $8N$ | - |
| 5 | FullyConn | $2N$ | $1\times1$ |

Table S2: Network architecture used for group discrimination network (Top NN). The output of the network is a vector of dimension 2, values corresponding to the probability that the data tuple belongs to particular class.

| Layer | Type | Maps and Neurons | Filter Size |
|---|---|---|---|
| 0 | Input | $16M \times 64 \times 7N$ | - |
| 1 | Flatten | $7168N$ | - |
| 2 | Dropout (p=0.5) | - | - |
| 3 | FullyConn | $8N$ | $1\times1$ |
| 4 | ReLU | $8N$ | - |
| 5 | FullyConn | $25N$ | $1\times1$ |

Table S3: Network architecture used for subject discrimination network (adversary). The output of the network is a vector of dimension 25, values corresponding to the probability that the data tuple belongs to particular subject.

## 1.2 Topographical maps

| Layer | Type | Maps and Neurons | Filter Size |
|---|---|---|---|
| 0 | Input | $3M \times 64 \times 64N$ | - |
| 1 | Conv | $16M \times 32 \times 32N$ | 5×5 |
| 2 | ReLU | $16M \times 32 \times 32N$ | - |
| 3 | MaxPool | $16M \times 16 \times 16N$ | 2×2 |
| 4 | Conv | $32M \times 16 \times 16N$ | 5×5 |
| 5 | ReLU | $32M \times 16 \times 16N$ | - |
| 6 | MaxPool | $32M \times 8 \times 8N$ | 2×2 |
| 7 | Conv | $64M \times 8 \times 8N$ | 3×3 |
| 8 | ReLU | $64M \times 8 \times 8N$ | - |
| 9 | MaxPool | $64M \times 4 \times 4N$ | 2×2 |

Table S4: Network architecture used for EEG feature extraction network (Base CNN). The output of the network is a tensor of dimensions $64 \times 4 \times 4$.

| Layer | Type | Maps and Neurons | Filter Size |
|---|---|---|---|
| 0 | Input | $64M \times 4 \times 4N$ | - |
| 1 | Flatten | 1024N | - |
| 2 | Dropout (p=0.5) | - | - |
| 3 | FullyConn | 8N | 1×1 |
| 4 | ReLU | 8N | - |
| 5 | FullyConn | 2N | 1×1 |

Table S5: Network architecture used for group discrimination network (Top NN). The output of the network is a vector of dimension 2, values corresponding to the probability that the data tuple belongs to particular class.

| Layer | Type | Maps and Neurons | Filter Size |
|---|---|---|---|
| 0 | Input | $64M \times 4 \times 4N$ | - |
| 1 | Flatten | 1024N | - |
| 2 | Dropout (p=0.5) | - | - |
| 3 | FullyConn | 8N | 1×1 |
| 4 | ReLU | 8N | - |
| 5 | FullyConn | 25N | 1×1 |

Table S6: Network architecture used for subject discrimination network (adversary). The output of the network is a vector of dimension 25, values corresponding to the probability that the data tuple belongs to particular subject.

## 2 Training curves

### 2.1 Time-Frequency Maps

| Hyperparameter | Value |
|---|---|
| Learning Rate | 0.001 |
| Learning Rate Decay | 0.0001 |
| Weight Decay | 0.001 |

Table S7: List of hyperparameters used for training the networks on TF maps.

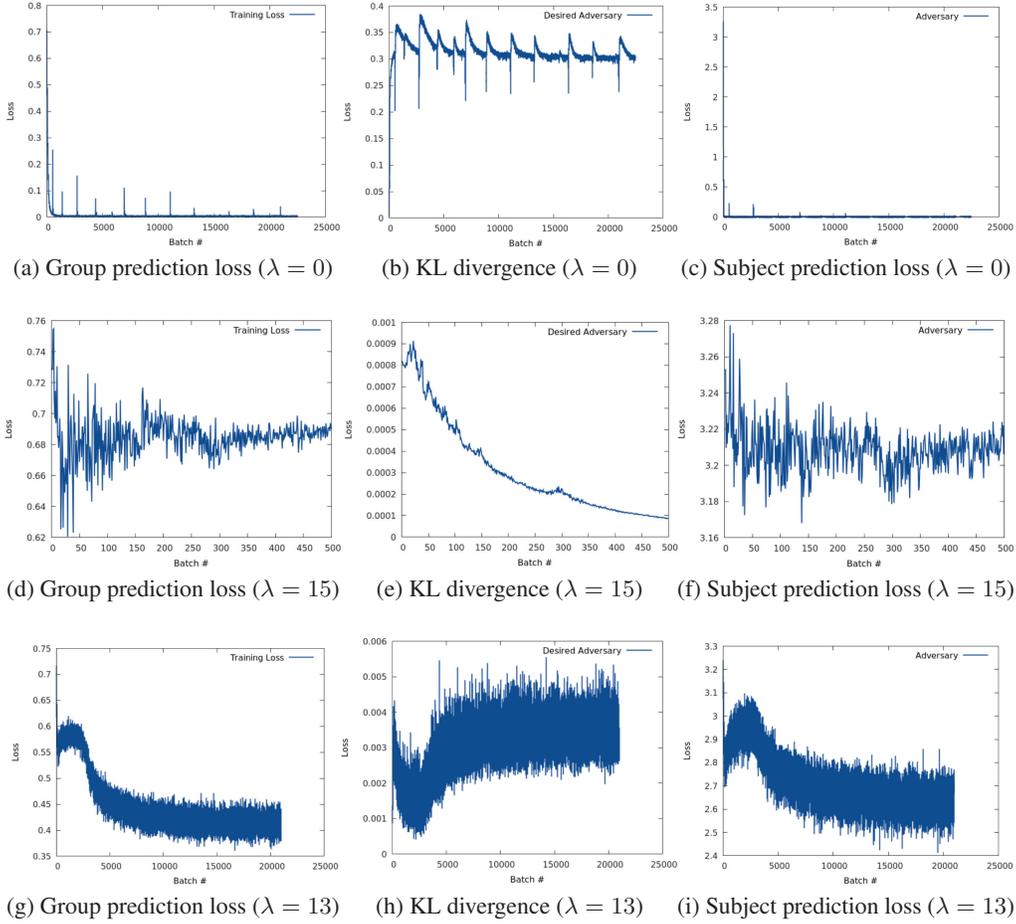

(a) Group prediction loss ($\lambda = 0$)  (b) KL divergence ($\lambda = 0$)  (c) Subject prediction loss ($\lambda = 0$)

(d) Group prediction loss ($\lambda = 15$)  (e) KL divergence ($\lambda = 15$)  (f) Subject prediction loss ($\lambda = 15$)

(g) Group prediction loss ($\lambda = 13$)  (h) KL divergence ($\lambda = 13$)  (i) Subject prediction loss ($\lambda = 13$)

Figure S2: Time-Frequency Maps Training curves for three different weight values to the subject predictor regularizer.

## 2.2 Topographical Maps

| Hyperparameter | Value |
| --- | --- |
| Learning Rate | 0.001 |
| Learning Rate Decay | 0.001 |
| Weight Decay | 0.03 |

Table S8: List of hyperparameters used for training the networks on Topographical maps.

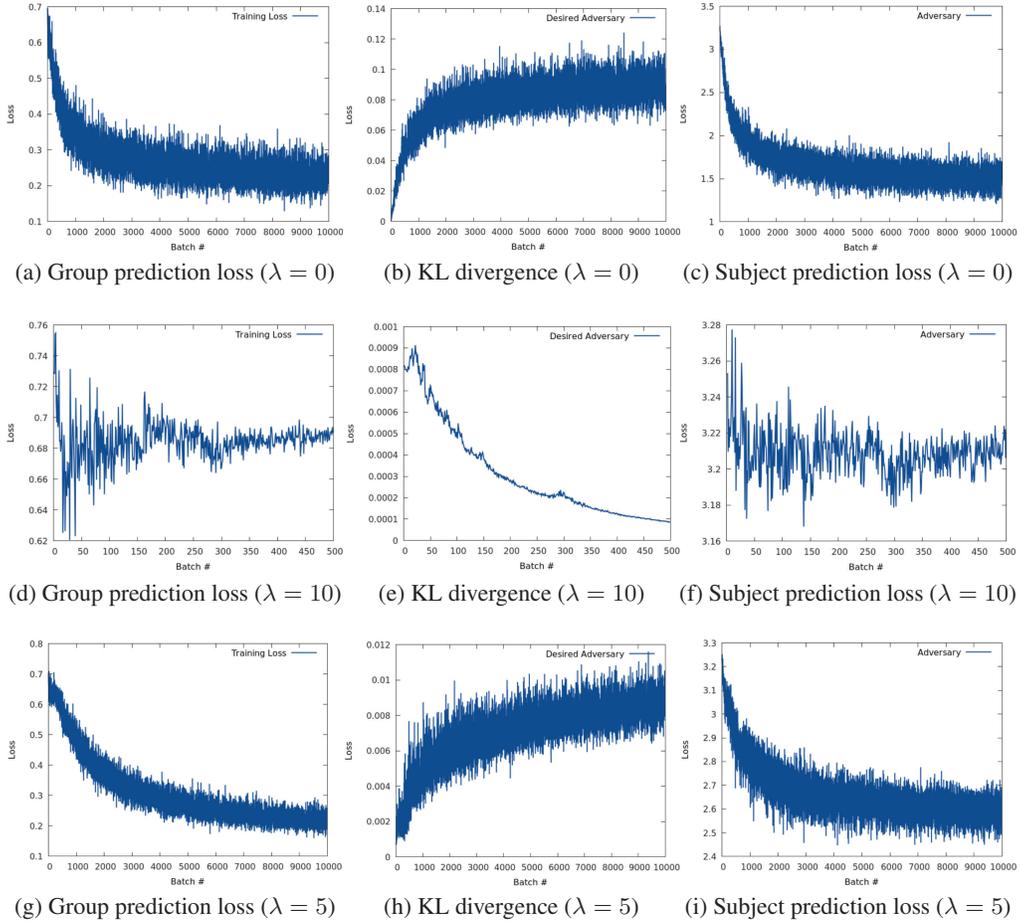

(a) Group prediction loss ($\lambda = 0$)  (b) KL divergence ($\lambda = 0$)  (c) Subject prediction loss ($\lambda = 0$)

(d) Group prediction loss ($\lambda = 10$)  (e) KL divergence ($\lambda = 10$)  (f) Subject prediction loss ($\lambda = 10$)

(g) Group prediction loss ($\lambda = 5$)  (h) KL divergence ($\lambda = 5$)  (i) Subject prediction loss ($\lambda = 5$)

Figure S3: Topographical Maps Training curves for three different weight values to the subject predictor regularizer.

## 3 ccCAM on Topographical Maps

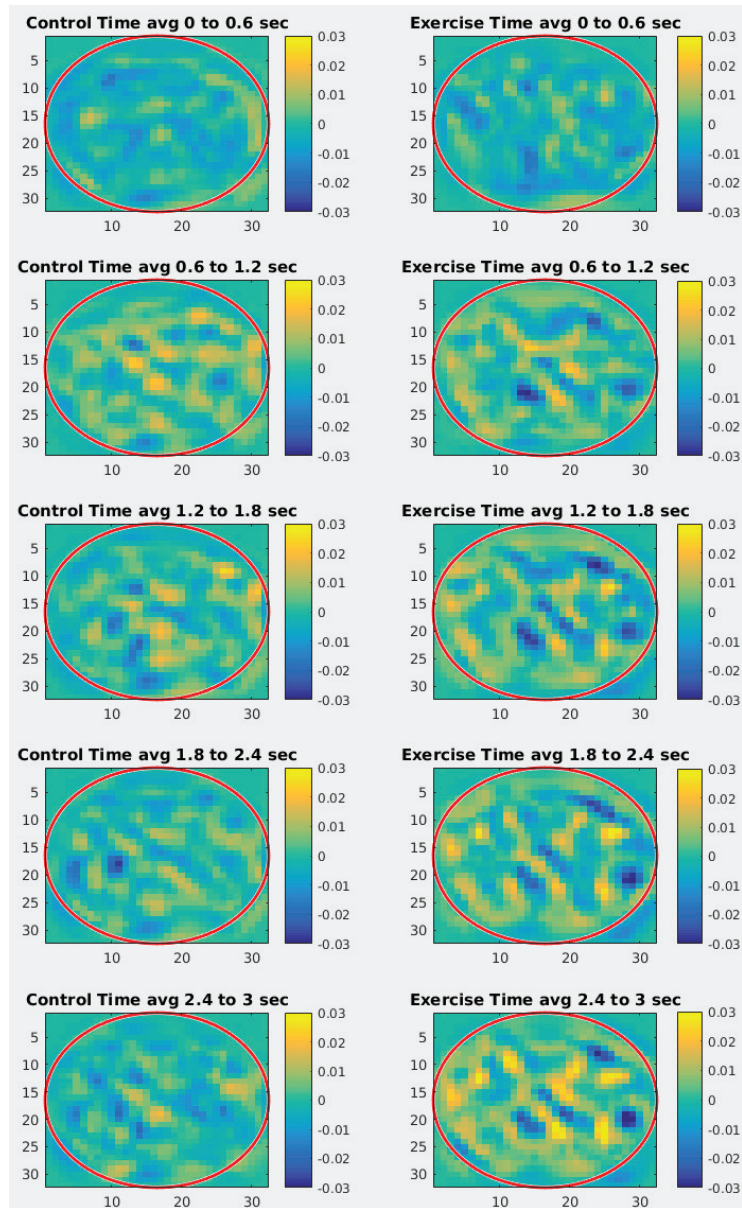

Figure S4: ccCAM matrices generated for Topographical Maps of 23–33 Hz activity group-averaged over time windows of 0.6 sec. The CAMs have higher values for EXE group as compared to the CON group which indicates that the network learns to look at the differences in EEG activity while performing the fixed-force handgrip task induced by acute bout of high-intensity exercise. The time described here is calculated from 0.5 sec after the presentation of visual cue.